\documentclass[geosciences,article,accept,moreauthors,pdftex,10pt,a4paper]{mdpi}

\usepackage{booktabs} 
\usepackage{multirow}
\usepackage{soul} 
\usepackage{microtype}
\usepackage{upgreek}
\usepackage{epstopdf}
\usepackage{amssymb}

\definecolor{red}{rgb}{1,0,0}

\firstpage{1} 
\pubvolume{8}
\issuenum{9}
\articlenumber{314}
\pubyear{2018}
\copyrightyear{2018}
\history{Received: 29 June 2018; Accepted: 20 August 2018; Published: 23 August 2018}

\updates{yes} 
 
\Title{The Critical Role of the Boundary Layer Thickness for the Initiation of Aeolian Sediment Transport}

\Author{Thomas P\"ahtz $^{1,2,}$*\orcidA{}, 
Manousos Valyrakis $^{3}$\orcidB{}, Xiao-Hu Zhao $^{4}$ and Zhen-Shan Li $^{5,6}$}

\AuthorNames{Thomas P\"ahtz, Manousos Valyrakis, Xiao-Hu Zhao and Zhen-Shan Li}

\address{%
$^{1}$ \quad Institute of Port, Coastal and Offshore Engineering, Ocean College, Zhejiang University, \newline  866 Yu Hang Tang Road,  Hangzhou 310058, China\\
$^{2}$ \quad State Key Laboratory of Satellite Ocean Environment Dynamics, Second Institute of Oceanography, \newline 36 North Baochu Road,  Hangzhou 310012, China\\
$^{3}$ \quad Infrastructure and Environment Research Division, School of Engineering, University of Glasgow, Glasgow G12 8QQ, 
UK; Manousos.Valyrakis@glasgow.ac.uk\\
$^{4}$ \quad Division of Environmental Management and Policy, School of Environment, Tsinghua University, Beijing 100084, China; 360129762@qq.com\\ 
$^{5}$ \quad College of Environmental Science and Engineering, Peking University, Beijing 100871, China; lizhenshan@pku.edu.cn\\
$^{6}$ \quad The Key Laboratory of Water and Sediment Sciences, Ministry of Education, Peking University, Beijing 100871,~China}

\corres{Correspondence: 0012136@zju.edu.cn}

\abstract{Here, we propose a conceptual framework of Aeolian sediment transport initiation that includes the role of turbulence. Upon increasing the wind shear stress $\tau$ above a threshold value $\tau^\prime_t$, particles resting at the bed surface begin to rock in their pockets because the largest turbulent fluctuations of the instantaneous wind velocity above its mean value $\overline{u}$ induce fluid torques that exceed resisting torques. Upon a slight further increase of $\tau$, rocking turns into a rolling regime (i.e.,~rolling threshold $\tau_t\simeq\tau^\prime_t$) provided that the ratio between the integral time scale $T_i\propto\delta/\overline{u}$ (where~$\delta$ is the boundary layer thickness) and the time $T_e\propto\sqrt{d/[(1-1/s)g]}$ required for entrainment (where $d$ is the particle diameter and $s$ the particle--air--density ratio) is sufficiently large. Rolling then evolves into mean-wind-sustained saltation transport provided that the mean wind is able to compensate energy losses from particle-bed rebounds. However, when $T_i/T_e$ is too small, the threshold ratio scales as $\tau_t/\tau^\prime_t\propto{T}_e/T_i\propto{s}d^2/\delta^2$, consistent with experiments. Because $\delta/d$ controls $T_i/T_e$ and the relative amplitude of turbulent wind velocity fluctuations, we qualitatively predict that Aeolian sediment transport in natural atmospheres can be initiated under weaker (potentially much weaker) winds than in wind tunnels, consistent with indirect observational evidence on Earth and Mars.}

\keyword{Aeolian sand transport; saltation; aerodynamic entrainment threshold; saltation threshold; initiation threshold; static threshold; dynamic threshold; Venus wind tunnel; Titan wind tunnel}

\begin{document}

\section{Introduction}
\begin{figure}[H]
\centering
\includegraphics[width=1.0\columnwidth]{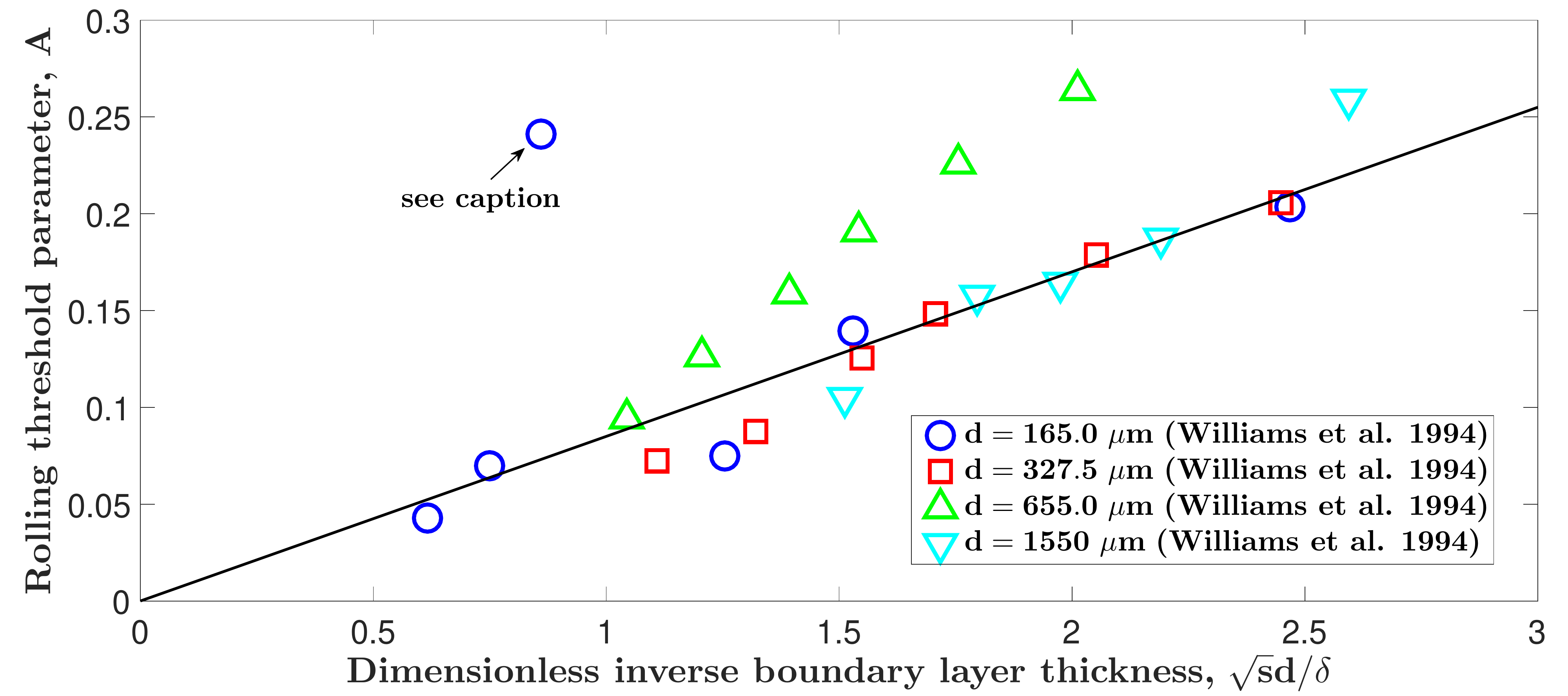}
\caption{Rolling threshold parameter $A=u_{\ast t}/\sqrt{(s-1)gd}$ versus dimensionless inverse boundary layer thickness $\sqrt{s}d/\delta$. Symbols correspond to the experimental data by Williams et al.~\cite{Williamsetal94}, who set up their wind tunnel in a manner that produces a developing turbulent boundary layer, for four different sediments consisting of nearly uniform, cohesionless particles. The line corresponds to the scaling Equation~(\ref{WilliamsScaling}). We suspect that the one extreme outlier for $d=165~\mu$m may either have been a faulty measurement or be associated with the observation that the boundary layer for this particular sand sample was not always fully turbulent~\cite{Williamsetal94}.}
\label{WilliamsExperiments}
\end{figure}

What is the wind shear stress ($\tau$) required to initiate Aeolian sediment transport by atmospheric wind on Earth and other planetary bodies? The answer to this question is thought to be critical for predicting dust aerosol emission in climate models~\cite{Koketal14a,Koketal14b,Hausteinetal15,Koketal18}, planetary sediment transport and bedform evolution~\cite{Bourkeetal10,Lorenz14,Rasmussenetal15}, and for inferring atmospheric wind conditions from sediment transport observations~\mbox{\cite{Bridgesetal12a,Ayoubetal14,LindhorstBetzler16}}. However, whether a critical fluid shear stress, and thus the mean turbulent flow, solely controls transport initiation is actually not so clear, as the wind tunnel experiments by \mbox{Williams et al.}~\cite{Williamsetal94} indicate. In contrast to most, if not all, other aerodynamic entrainment experiments in the literature, these authors intentionally set up their wind tunnel in a manner that produces a developing turbulent boundary layer rather than a fully developed one. This setup allowed them to study the potential influence of both the mean turbulent flow and turbulent fluctuations on transport initiation, which change quite significantly in a developing boundary layer. For example, while the boundary layer thickness $\delta$ (i.e., the distance from the bed surface to the elevation at which the local mean wind velocity is equal to $99\%$ of the free stream velocity) and turbulent kinetic energy increase downwind, the wind shear velocity $u_\ast=\sqrt{\tau/\rho_a}$, where $\rho_a$ is the air density, decreases. In~fact, for all of their four tested sediments (nearly uniform, cohesionless), Williams et al.~\cite{Williamsetal94} measured that, upon~increasing the free stream wind velocity, particles are entrained first at the downwind end of the test section despite the fact that $u_\ast$ is smallest there, whereas entrainment at the upwind end required much larger free stream wind velocities (a similar observation was also made by Bagnold~\cite{Bagnold41}). Furthermore, we show in Figure~\ref{WilliamsExperiments} that most of these authors' measurements of $u_{\ast t}$ (the threshold value of $u_\ast$ at which particles begin to roll, also called `detachment threshold'~\cite{DeVetetal14}) roughly collapse following the scaling
\begin{eqnarray}
 A\equiv\frac{u_{\ast t}}{\sqrt{(s-1)gd}}\propto\frac{\sqrt{s}d}{\delta}, \label{WilliamsScaling}
\end{eqnarray}
where $A$ is the rolling threshold parameter (the square-root of the threshold Shields number), $s=\rho_p/\rho_a$ the particle--air--density ratio, $g$ the gravitational constant, and $d$ the characteristic particle diameter.
This scaling is very unusual as it fundamentally differs from the currently well-established point of view~\cite{Iversenetal87,Burretal15} that, for cohesionless sediments, $A$ is only a function of $s$ and the particle Reynolds number $\mathrm{Re}_d=u_{\ast t}d/\nu$, or equivalently of the Galileo number $\mathrm{Ga}=\sqrt{(s-1)gd^3}/\nu=\mathrm{Re}_d/A$ (the use of which is preferential as it does not incorporate a dependency on $u_{\ast t}$), where $\nu$ is the kinematic air viscosity. Figure~\ref{WilliamsExperiments} also shows that $A$ varies by a factor of about $3$ for a given $d$, which is much beyond the usual variability of existing threshold measurements for fully developed turbulent boundary layers~\cite{Raffaeleetal16}. Clearly, the data by Williams et al.~\cite{Williamsetal94}, though obtained for the rather unusual case of a developing turbulent boundary layer, require an explanation, or, at the very least, raise important questions that need to be addressed: what are the roles that turbulent fluctuations play in Aeolian sediment transport initiation, and what makes the scaling of $A$ for developing boundary layers so fundamentally different from fully developed ones? In addition, how much can we rely on theoretical predictions adjusted to wind tunnel experiments given that turbulence in natural atmospheres has turbulent fluctuation properties that are fundamentally different from that of wind tunnels because of a much thicker boundary layer ($\delta\sim$0.1--10~km on Venus, Earth, Mars, and Titan~\cite{Lorenzetal10,Petrosyanetal11,Koketal12,Lebonnoisetal18})? The~latter question is of particular relevance because the existence of gravel megaripples (e.g., in the Puna of Argentina~\cite{DeSilvaetal13}) suggests that even centimeter-sized particles may be susceptible to aerodynamic entrainment by atmospheric wind speeds that, according to wind tunnel experiments, should be way too low to have an effect. Likewise, based~on wind tunnel experiments, it is difficult to explain how sand transport can be initiated in the present climate on Mars~\cite{SullivanKok17}, which is problematic because observations suggest widespread and persistent sediment activity~\cite{Bridgesetal12a,Bridgesetal12b,Silvestroetal13,Chojnackietal15}, even of very coarse sand~\cite{Bakeretal18}.

Here, we tackle these questions by developing, step by step, a conceptual framework of the initiation of cohesionless sediment transport by wind (Section~\ref{Framework}). Cohesive effects, which become significant for $d\lesssim200~\upmu$m and crucial for $d\lesssim100~\upmu$m~\cite{GreeleyIversen85,ShaoLu00,Merrisonetal07,Merrison12,Koketal12}, are neglected to put more emphasis on this framework's novelty but can be relatively easily incorporated in more refined frameworks in future studies. In fact, the most important novel aspect of our framework is that we consider the critical roles of the amplitude and duration of turbulent velocity fluctuations~\cite{Diplasetal08,Valyrakisetal10} using the energy criterion of entrainment~\cite{Valyrakisetal13}, which is well established in the fluvial sediment transport community, but largely unknown in the Aeolian sediment transport community and among planetary scientists. We~test the predictions of our framework with existing and new threshold measurements that have been carried out for varying density ratio $s$, Galileo number $\mathrm{Ga}$ 
, and boundary layer thickness $\delta$ (Section~\ref{Experiments}), and~discuss the potential threshold dependency on $s$ for a constant $\mathrm{Ga}$ (Section~\ref{Discussion}). This~dependency is quite controversial~\cite{Iversenetal87,Luetal05,Burretal15}, and we suggest that it may be an artifact of a strongly varying thickness $\delta$ of the boundary layers produced by the different experimental facilities.

\section{Conceptual Framework of Aeolian Sediment Transport Initiation} \label{Framework}
\unskip
\subsection{Torque Balance Criterion Associated with a Turbulent Fluctuation Event} \label{FluctuationTorque}

What are the minimal requirements for the entrainment of a test particle resting at the bed surface by a turbulent flow? First, as entrainment in a rolling motion is usually the easiest~\cite{Valyrakisetal13}, it is required that the largest instantaneous fluid torques acting on this particle exceed resisting torques. Such~torques are induced by the largest positive fluctuations of the instantaneous fluid velocity $u$ (evaluated at an effective particle elevation $z_{\mathrm{eff}}$) around its mean value $\overline{u}$. Numerous theoretical studies have derived expressions for this torque balance criterion (e.g., see the review by Dey and Ali~\cite{DeyAli18}). For~cohesionless particles, one can summarize such expressions in the functional form
\begin{eqnarray}
 \frac{u_m}{\sqrt{(s-1)gd}}\equiv\alpha\frac{\overline{u}}{\sqrt{(s-1)gd}}\geq f_1(G), \label{umean}
\end{eqnarray}
where $G$ is a bed geometry parameter, which takes into account the size distribution, arrangement, and~shape of bed particles; $u_m$ is the characteristic value of $u$ associated with its largest positive fluctuations and $\alpha=u_m/\overline{u}$ a parameter that characterizes the maximal relative amplitude of fluctuations. There~are further dependencies on the drag and lift coefficients, both of which we assume to be constant for simplicity. The threshold parameter $A^\prime=u^\prime_{\ast t}/\sqrt{(s-1)gd}$ associated with the fluctuation-induced torque equilibrium now follows from Equation~(\ref{umean}) and the fact that the ratio $u^\prime_{\ast t}/\overline{u}$ is predominantly a function of the particle Reynolds number $\mathrm{Re}^\prime_d=\mathrm{Ga}A^\prime$ (which controls the flow velocity profile) and the bed geometry $G$ (which controls $z_{\mathrm{eff}}/d$):
\begin{eqnarray}
 A^\prime=f_2(G,\mathrm{Ga})\frac{\min(\overline{u})}{\sqrt{(s-1)gd}}=\alpha^{-1}f_1(G)f_2(G,\mathrm{Ga}). \label{Aprime}
\end{eqnarray}

At this point, we would like to emphasize that a torque balance criterion does not result in a dependency of $A^\prime$ on the density ratio $s$ for a constant Galileo number $\mathrm{Ga}$. For this reason, it has been argued that the additional dependency of Aeolian sediment transport initiation on $s$ suggested by some experiments~\cite{Iversenetal87,Burretal15} is a signature of a particle--inertia effect, such as a dynamic collision force caused by the impact of a hopping particle with the test particle~\cite{Iversenetal87}. However, this impact--force hypothesis cannot explain why the initiation of saltation transport (i.e., the transport regime in which particles hop along the bed in large ballistic trajectories) is usually preceded by a rolling regime further upwind~\cite{Bagnold41,Iversenetal87}. In contrast, a strong density ratio dependency is inherent to dynamic saltation threshold models~\cite{ClaudinAndreotti06,Kok10b,Pahtzetal12,Berzietal17,PahtzDuran18a}, some of which simultaneously explain available measurements of Aeolian and viscous and turbulent fluvial sediment transport thresholds despite not being fitted to these measurements~\cite{PahtzDuran18a}.

\subsection{Energy Criterion Associated with a Turbulent Fluctuation Event}
When the threshold parameter $A^\prime$ is exceeded, the test particle will begin to roll whenever the largest turbulent fluctuation events occur. However, in order for this particle to leave its pocket and be truly entrained, such events must last sufficiently long~\cite{Diplasetal08,Valyrakisetal10,Valyrakisetal13}. Otherwise, it will just roll back and forth around its pocket, which is visually perceived as an oscillatory rocking motion. Note~that this rocking regime has not only been observed in fluvial environments~\cite{Diplasetal08}, but also in Aeolian environments~\mbox{\cite{GreeleyMarshall85,Burretal15}}. A criterion that roughly quantifies whether actual entrainment, beyond rocking, occurs was first given by Diplas et al.~\cite{Diplasetal08} (the `impulse criterion'~\cite{Valyrakisetal10}) and later refined by \mbox{Valyrakis et al.}~\cite{Valyrakisetal13}: (the `energy criterion', which shares some similarities with energy accumulation-based models of the resuspension of aerosols in turbulent flows~\cite{Ziskindetal95,ReeksHall01}). According to the energy criterion, entrainment occurs if
\begin{eqnarray}
 \langle u\rangle^2\langle v\rangle T_i\geq c_T(G)f_1(G)^2(s-1)gd^2 \label{EnergyCriterion}
\end{eqnarray}
during a turbulent fluctuation event that exceeds the torque equilibrium [Equation~(\ref{umean})], where $\langle v\rangle$ and $\langle u\rangle$ are the particle and effective instantaneous fluid velocity, respectively, averaged over the duration $T_i$ of this event. As we are interested in the minimal requirement for entrainment, we~take $T_i$ as proportional to the integral time scale, which is a measure for the maximal temporal correlation of fluid velocity fluctuations~\cite{VarshneyPoddar11}, and $\langle u\rangle$ as proportional to the characteristic fluid velocity $u_m$ resulting from the largest positive fluctuations. However, we replace these proportionalities by equalities (i.e., $\langle u\rangle=u_m$) and incorporate the associated constant proportionality factors in the bed-geometry-dependent function $c_T(G)$. Importantly, for mild bed slopes, $c_T(G)$ changes with the average pivoting angle $\theta$ (i.e., the angle between the horizontal and the lever arm through the pivoting point) as $c_T(G)\propto\tan\theta(1-\sin\theta)$~\cite{Valyrakisetal13} and thus decreases with $\theta$ for $\theta>38^\circ$. Because, for nearly uniform sediments, typical pivoting angles are significantly larger (e.g., $\theta=60^\circ$ for a two-dimensional close packing), we consider $c_T(G)$ as a function that decreases with particle exposure to the flow. Using $\langle u\rangle=u_m$ and Equation~(\ref{umean}) and defining the entrainment time scale $T_e=d/\langle v\rangle$, we can rewrite Equation~(\ref{EnergyCriterion}) as
\begin{eqnarray}
 T_i/T_e\geq c_T(G). \label{TiTe}
\end{eqnarray}

Equation~(\ref{TiTe}) means that, if the threshold parameter $A^\prime$ is exceeded (i.e., the largest turbulent fluctuations exceed the torque balance), the ratio $T_i/T_e$ must exceed the critical value $c_T(G)$ in order for the test particle to roll out of its pocket after a finite period of time. Hence, the rolling threshold parameter $A$ is given by
\begin{eqnarray}
 A\simeq A^\prime=\alpha^{-1}f_1(G)f_2(G,\mathrm{Ga})\quad\text{if}\quad T_i/T_e\geq c_T(G). \label{Afluc}
\end{eqnarray}

\subsection{Torque Balance Criterion Associated with the Mean Turbulent Flow}
Even if the typical duration of turbulent velocity fluctuations is much too short [$T_i/T_e\ll c_T(G)$], entrainment can still occur when the mean turbulent flow is so strong that the mean fluid torques acting on the test particle exceed resisting torques. Instead of Equations~(\ref{umean}) and (\ref{TiTe}), the entrainment condition then becomes
\begin{eqnarray}
 \overline{u}/\sqrt{(s-1)gd}\geq f_1(G). \label{umean2}
\end{eqnarray}

Analogous to Equation~(\ref{Aprime}), the rolling threshold parameter $A$ associated with an arbitrary entrainment condition, such as Equation~(\ref{umean2}), can be expressed as
\begin{eqnarray}
 \frac{A}{A^\prime}=\frac{\alpha}{f_1(G)}\;\frac{\min(\overline{u})}{\sqrt{(s-1)gd}}, \label{A}
\end{eqnarray}
where we used Equations~(\ref{umean}) and (\ref{Aprime}). Hence, inserting Equation~(\ref{umean2}) in Equation~(\ref{A}) yields
\begin{eqnarray}
 A=\alpha A^\prime=f_1(G)f_2(G,\mathrm{Ga})\quad\text{if}\quad T_i/T_e\ll c_T(G). \label{Amean}
\end{eqnarray}

\subsection{The Intermediate Regime between Mean Flow Entrainment and Fluctuation-Induced Entrainment}
We now consider the intermediate case in which $T_i/T_e$ is smaller than $c_T(G)$, but still so large that mean flow entrainment is not yet required (i.e., neither Equation~(\ref{umean}) nor Equation~(\ref{umean2}) are characterizing entrainment), which implies that the energy criterion Equation~(\ref{EnergyCriterion}) is solely controlling entrainment. Using again $\langle u\rangle=u_m$ and $T_e=d/\langle v\rangle$, we rewrite Equation~(\ref{EnergyCriterion}) as
\begin{eqnarray}
 \frac{\overline{u}^2}{(s-1)gd}\geq c_T(G)\alpha^{-2}f_1(G)^2\frac{T_e}{T_i}. \label{EnergyCriterion2}
\end{eqnarray}

Hence, inserting Equation~(\ref{EnergyCriterion2}) in Equation~(\ref{A}) yields
\begin{eqnarray}
 \text{Intermediate regime:}\quad\frac{A^2}{A^{\prime 2}}\equiv\frac{\tau_t}{\tau^\prime_t}=c_T(G)\frac{T_e}{T_i}. \label{Ain}
\end{eqnarray}

\subsection{The Time Scale Ratio $T_e/T_i$}
For a cohesionless test particle at the torque balance equilibrium, the torques induced by fluid drag and lift just overcome the resisting torque induced by its submerged weight~\cite{DeyAli18}. Once this particle begins to roll within its pocket, these torques change because of increasing pivoting angle. Hence, if the fluid drag and lift force remain approximately the same, this particle will be accelerated at a rate proportional to $g(1-1/s)$ that depends only on the pivoting angle, which is why we estimate the entrainment time scale as
\begin{eqnarray}
 T_e=f_3(G)\sqrt{d/[(1-1/s)g]},
\end{eqnarray}
where $f_3(G)$ is another bed-geometry-dependent coefficient. Using that the integral length scale $L_i=T_i\overline{u}$~\cite{VarshneyPoddar11}, which is a measure for the maximal spatial correlation of fluid velocity fluctuations, is~approximately proportional to the boundary layer thickness $\delta$ ($L_i\approx0.4\delta$~\cite{AlhamdiBailey17}), we thus obtain
\begin{eqnarray}
 \frac{T_i}{T_e}\approx0.4f_3(G)^{-1}\frac{\sqrt{(s-1)gd}}{\overline{u}}\;\frac{\delta}{\sqrt{s}d}. \label{TimeScaleRatio}
\end{eqnarray}

When $\overline{u}$ is just large enough to exceed the respective threshold criterion (Equation~(\ref{umean}) or Equation~(\ref{umean2}) or Equation~(\ref{EnergyCriterion2})), we can insert Equation~(\ref{A}) in Equation~(\ref{TimeScaleRatio}), which yields
\begin{eqnarray}
 c_T(G)^{-1}\frac{T_i}{T_e}\approx \alpha f_T(G)^{-1}\frac{A^\prime}{A}\;\frac{\delta}{\sqrt{s}d}, \label{TimeScaleRatio2}
\end{eqnarray}
where $f_T(G)=2.5c_T(G)f_1(G)f_3(G)$ is a bed-geometry-dependent coefficient. Importantly, the smaller the $f_T(G)$, the more the bed particles are exposed to the flow because particle exposure decreases $c_T(G)$ (as discussed before), the flow velocity needed to exceed the torque equilibrium (and thus $f_1(G)$), and the time needed for entrainment (and thus $f_3(G)$). For the intermediate regime, Equation~(\ref{TimeScaleRatio2}) approximately implies
\begin{eqnarray}
 \text{Intermediate regime:}\quad\frac{A}{A\prime}\approx\alpha^{-1}f_T(G)\frac{\sqrt{s}d}{\delta}, \label{TimeScaleRatio3}
\end{eqnarray}
where we used Equation~(\ref{Ain}). Equation~(\ref{TimeScaleRatio3}) resembles the scaling (Equation~(\ref{WilliamsScaling})) that collapses most of the experimental data by Williams et al.~\cite{Williamsetal94} because $A^\prime$ is relatively constant for the experimental range of Galileo numbers ($\mathrm{Ga}\in(19,~554)$). This constancy of $A^\prime$ follows from the constancy of the saltation threshold $A^s$ measured in wind tunnel experiments that mimic the atmosphere on Earth and produce a fully developed turbulent boundary layer (Figure~15a of Dur\'an et al.~\cite{Duranetal11}), for which $A^s\simeq A\simeq A^\prime$ ~\cite{Iversenetal87} (reminder: this equality does not apply to the developing boundary layer wind tunnel experiments by Williams et al.~\cite{Williamsetal94}). Finally, combining Equation~(\ref{TimeScaleRatio3}) with Equations~(\ref{Afluc}) and~(\ref{Amean}), we obtain
\begin{eqnarray}
 \frac{A}{A\prime}&\approx& 
 \begin{matrix}
 1 & \text{if} & B<1 & \text{(Fluctuation-torque-balance regime)} \\
 B & \text{if} & 1\leq B\leq\alpha & \text{(Intermediate regime)} \\
 \alpha & \text{if} & B>\alpha & \text{(Mean-torque-balance regime)} \\
 \end{matrix}
 \label{TimeScaleRatio4} \\
 B&=&\alpha^{-1}f_T(G)\frac{\sqrt{s}d}{\delta}. \nonumber
\end{eqnarray}

However, note that the actual transitions between the different regimes are likely not as sharp as predicted by our very simplified framework.

\subsection{The Maximal Relative Amplitude of Turbulent Velocity Fluctuations \label{alpha} } 
The maximal relative amplitude of turbulent velocity fluctuations is characterized by the parameter $\alpha=u_m/\overline{u}$, which depends on the characteristic value $u_m$ of the effective instantaneous fluid velocity $u$ associated with its largest positive fluctuations. This value is controlled by the distribution of $u/u_\ast$, which has been extensively studied in the past (e.g., see the review by Smits et al.~\cite{Smitsetal11}). It has been found that this distribution is predominantly controlled by the dimensionless distance from the wall ($zu_\ast/\nu$) and the boundary layer Reynolds number $\delta u_\ast/\nu$. Hence, as $u$ is evaluated at the effective particle elevation $z_{\mathrm{eff}}$ (which slightly depends on $G$), $\alpha$ obeys the functional form
\begin{eqnarray}
 \alpha=f_\alpha(G,\mathrm{Ga},\delta/d).
\end{eqnarray}

In particular, $\alpha$ exhibits a slight log-like increase with $\delta/d$ for a constant $\mathrm{Ga}$~\cite{MarusicKunkel03}, which was used by Lu et al.~\cite{Luetal05} to explain why $A$ is smaller in air than in water. According to Equation~(\ref{TimeScaleRatio4}), $\alpha$ controls the maximal variability of $A$. The variability of the data by Williams et al.~\cite{Williamsetal94} therefore suggests that $\alpha$ may be significantly larger than $3$ for their wind tunnel experiments mimicking atmospheric conditions on Earth (Figure~\ref{WilliamsExperiments}), and thus even larger in wind tunnels that mimic such atmospheres and produce a fully developed boundary layer (larger $\delta$), and thus even larger in the field (much larger $\delta$), which~emphasizes the critical role of turbulent velocity fluctuations for the initiation of Aeolian sediment transport.

\subsection{The Rolling-Saltation Transition} \label{RollingSaltation}
Once particles begin to roll along the bed surface, they will be accelerated towards the mean near-bed fluid velocity, provided that the acceleration during fluctuation events exceeds the potential deceleration when such events are over (we expect that $T_i/T_e\geq c_T(G)$ is a sufficient condition fulfilling this requirement), and thus eventually begin to saltate (some particles may also be entrained directly into saltation~\cite{Valyrakisetal10,Valyrakisetal13}). However, whether this saltation motion is limited to the immediate vicinity of the bed or whether particles accumulate more and more height with each rebound at the bed surface depends on the ability of the flow to compensate the energy losses from such rebounds. P\"ahtz \& Dur\'an~\cite{PahtzDuran18a} recently showed that the dynamic-saltation-threshold parameter $A^r$ is associated with the question of whether the mean flow is able to compensate such losses for fully developed saltation transport (i.e., large particle hops), which is easier than compensating energy losses of particles hopping near the bed because the time between successive particle-bed rebounds is larger, and because the wind velocity increases with elevation. We thus hypothesize that the threshold parameter $A^s$ associated with the initiation of fully developed saltation obeys
\begin{eqnarray}
 A^s\simeq A\simeq A^\prime\quad\text{if}\quad T_i/T_e\geq c_T(G)\quad\text{AND}\quad A^\prime/A^r\geq C(\mathrm{Ga},s),
\end{eqnarray}
where $C(\mathrm{Ga},s)\geq1$ is a critical value of $A^\prime/A^r$ that depends on the energy gap between particle trajectories associated with fully developed saltation and particle trajectories associated with saltation near the bed surface, which depends on the density ratio $s$ and Galileo number $\mathrm{Ga}$~\cite{PahtzDuran18a}. The fact that $A^s$ and $A^r$ are usually close to each other~\cite{Bagnold41,Chepil45,MartinKok18a} (except for very large $s$, like on Mars~\cite{ClaudinAndreotti06,Almeidaetal08,Kok10a,Kok10b,Pahtzetal12}) implies that $C$ is usually close to unity (except for very large $s$).

\section{Test of Entrainment Framework with Existing and New Experimental Data} \label{Experiments}
The prediction of Equation~(\ref{TimeScaleRatio4}) for the intermediate regime has already been tested against the experimental data by Williams et al.~\cite{Williamsetal94} (Figure~\ref{WilliamsExperiments}). Though this prediction is, in general, consistent with these data, there are significant deviations for the intermediate particle diameter $d=655\;\upmu$m, for which we have no explanation. In order to further test Equation~(\ref{TimeScaleRatio4}), we scanned the literature for simultaneous rocking, rolling, and/or saltation threshold measurements. However, we would like to emphasize that, for conditions in which these thresholds significantly differ from each other, measurements of $A^\prime$ likely overestimate the actual rocking threshold as too short-lived rocking events may not be detected by the experimental setups. We therefore focus on the validation of the qualitative predictions of Equation~(\ref{TimeScaleRatio4}): $A\simeq A^\prime$ and thus likely $A^s\simeq A^\prime$ (from Section~\ref{RollingSaltation} and the fact that usually $A^s$ is significantly larger than the dynamic threshold $A^r$ in wind tunnels) if a critical value of the dimensionless boundary layer thickness $\delta/(\sqrt{s}d)$ is exceeded.

Simultaneous rocking, rolling, and saltation threshold measurements were carried out by \mbox{Iversen et al.}~\cite{Iversenetal76,IversenWhite82} in a large wind tunnel ($\delta\approx1.2~$m) that mimics atmospheric conditions on Earth~\cite{Iversenetal76,IversenWhite82}. Indeed, it was mentioned in a later study that there was nearly no difference between $A^s$, $A$, and $A^\prime$ for these experiments~\cite{Iversenetal87}. Likewise, Greeley and Marshall~\cite{GreeleyMarshall85} reported simultaneous measurements of $A^s$, $A$, and $A^\prime$ for their experiments under conditions mimicking the atmosphere on Venus in a pressurized wind tunnel, the so-called `Venus Wind Tunnel'. Similarly, Burr et al.~\cite{Burretal15} measured $A^s$ and $A^\prime$ under conditions mimicking the atmosphere on Titan using the same facility (renamed to `Titan Wind Tunnel'). Further saltation threshold measurements using this facility were reported by Greeley et al.~\cite{Greeleyetal84}. The Venus (Titan) Wind Tunnel produces a relatively thin boundary layer: Burr et al.~\cite{Burretal15} reported exactly one measurement ($\delta=1.9~$cm), which we therefore assume to be the value of $\delta$ for all threshold measurements carried out using this facility. Note that experimental wind tunnel studies with a non-standard and/or unclear experimental setup are not considered here. For example, de Vet et al.~\cite{DeVetetal14}, who also reported differences between rolling and saltation thresholds, placed patches of sand on a surface with a similar but not equivalent roughness rather than using a spatially homogeneous sand bed, which may have caused significant disturbances of the boundary layer. Likewise, in contrast to other experimental studies, these authors may not have manipulated the boundary layer at the wind tunnel entrance to create a fully developed boundary layer, in which case the boundary layer would have been a developing one (i.e., $\delta$ cannot be easily inferred).

\begin{figure}[H]
\centering
\includegraphics[width=0.98\columnwidth]{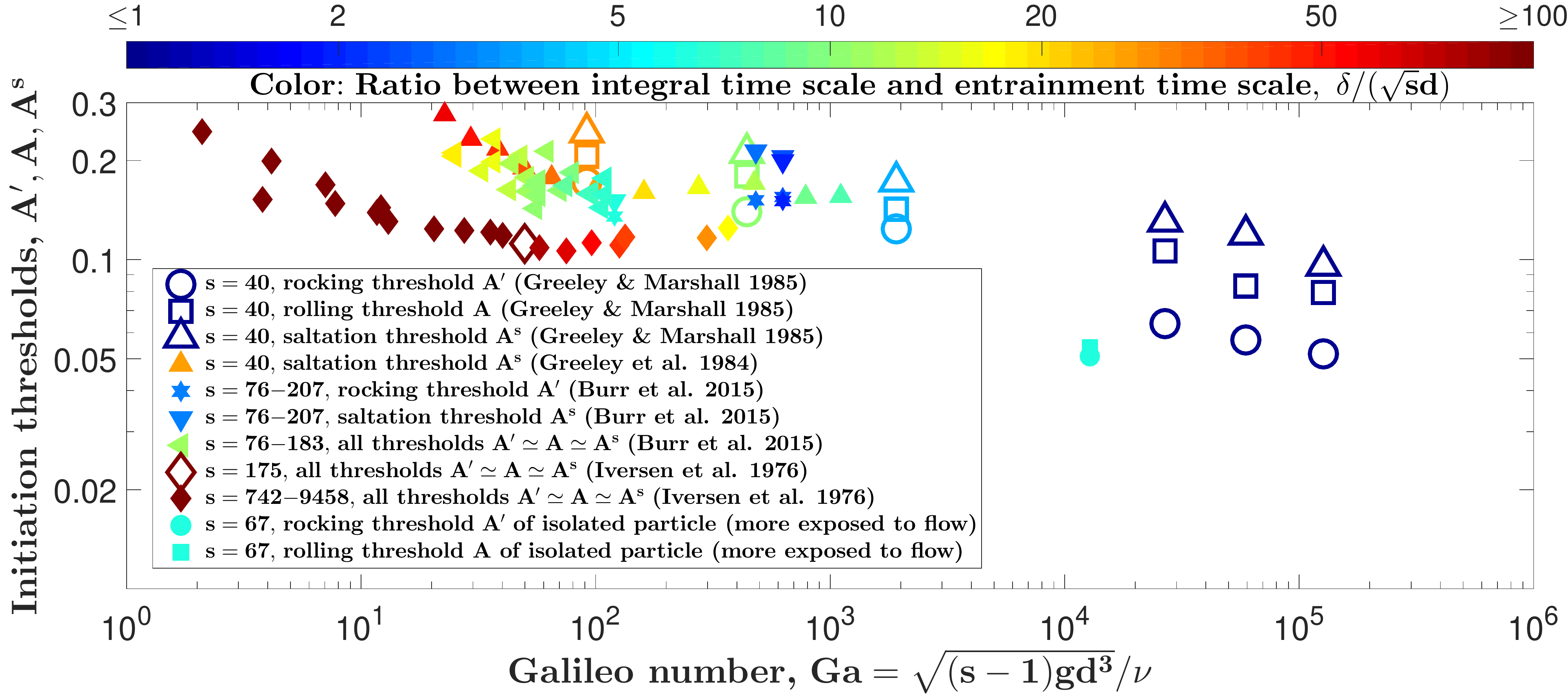}
\caption{Measurements of the rocking ($A^\prime$), rolling ($A$), and saltation ($A^s$) threshold parameters from the literature~\cite{Iversenetal76,Greeleyetal84,GreeleyMarshall85,Burretal15} and ourselves (isolated-particle measurements, see Figure~\ref{IsolatedParticleMeasurements}) versus the Galileo number $\mathrm{Ga}$. The color indicates the value of the dimensionless boundary layer thickness $\delta/(\sqrt{s}d)$, which is a measure for the ratio between the integral time scale $T_i$ and the entrainment time scale $T_e$.}
\label{ThresholdMeasurements}
\end{figure}

Figure~\ref{ThresholdMeasurements} shows the various measurements in log-scale, to emphasize the relative threshold gaps, where the estimated value of the dimensionless boundary layer thickness $\delta/(\sqrt{s}d)$ has been color-coded.
It can be seen that, for most conditions, saltation, rolling, and rocking thresholds become indistinguishable when $\delta/(\sqrt{s}d)$ exceeds a critical value [$\delta/(\sqrt{s}d)\gtrsim6.6$], in agreement with the theoretical prediction (Equation~(\ref{TimeScaleRatio4})). The only exception are the measurements by Greeley and Marshall~\cite{GreeleyMarshall85}, which exhibit significant relative threshold gaps even for their smaller sands ($d=105\;\upmu$m and $\delta/(\sqrt{s}d)\simeq29$). These measurements are particularly odd because Greeley et al.~\cite{Greeleyetal84} explicitly stated for the very same experiments that there was nearly no difference between $A^s$ and $A^\prime$ for small sands. Consistently, the saltation threshold measurements reported by Greeley et al.~\cite{Greeleyetal84} are below the saltation threshold measurements by Greeley and Marshall~\cite{GreeleyMarshall85} and rather close to the latter authors' rocking threshold measurements, as one would expect if there was no significant difference between rocking and saltation thresholds. Note that in a later study that compiles existing threshold data to demonstrate the effect of the density ratio $s$ on the initiation of fully developed saltation transport~\cite{Iversenetal87}, Greeley, Marshall, and coauthors do not incorporate the measurements of Ref.~\cite{GreeleyMarshall85} in their threshold diagrams, which suggests that there may have been a problem with these measurements. For these reasons, we do not consider these measurements as a contradiction to the existence of a critical value of $\delta/(\sqrt{s}d)$ above which rocking, rolling, and saltation thresholds become indistinguishable.

\begin{figure}[H]
\centering
\includegraphics[width=1.0\columnwidth]{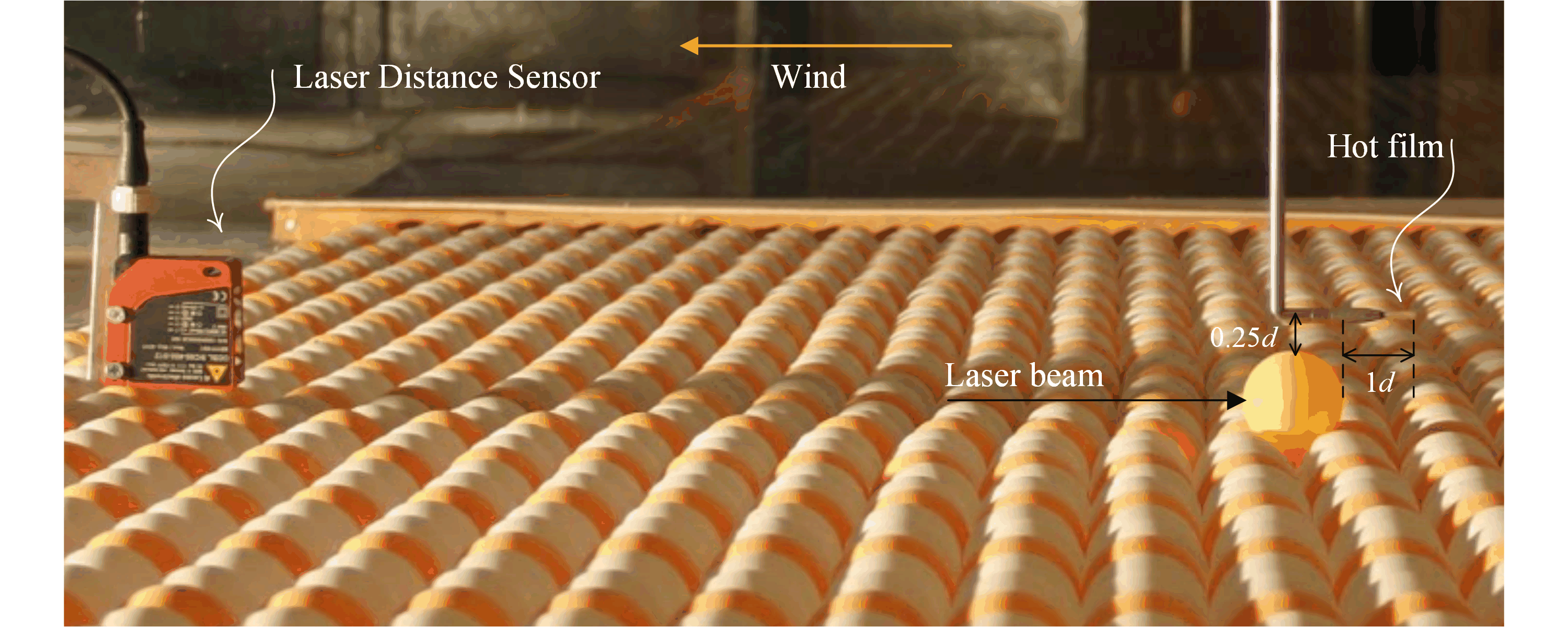}
\caption{Photograph of experimental setup. A $0.1~$mm resolution laser distance sensor (LDS, manufactured by Leuze Electronics, Owen, Germany) was used to track the motion of the target particle. A~red laser beam with $655~$nm wavelength launched by LDS shone at the center of the target particle. Once the particle began to rock along or against the wind direction, the corresponding displacement could be measured by the sensor. The measuring range of the sensor was set from the rest position of the target particle to the top of the downstream supporting particles, where the mobile particle would be considered to have been completely entrained. Calibration of the setup showed that the measured displacement of the target particle is a linear function of the signal intensity of the sensor.}
\label{IsolatedParticleMeasurements}
\end{figure}

Another qualitative prediction that can be tested is the effect of particle exposure. Equation~(\ref{TimeScaleRatio4}) predicts that the smaller the relative gap between the rolling threshold $A$ and rocking threshold $A^\prime$, the more exposed the bed particles are to the flow because of smaller $f_T(G)$. A large particle exposure can be achieved when placing a test particle on top of a close packing~\cite{FentonAbbott77}. We thus prepared a close packed bed ($4~\mathrm{m}\times2.4~\mathrm{m}$, $22.6~$m from the tunnel entrance) of uniform particles ($\rho_p=80.8~$kg/m$^3$, $d=40~$mm) in a wind tunnel at Beijing University ($\delta\approx2~$m) and placed an isolated particle of the same weight and size on its top (Figure~\ref{IsolatedParticleMeasurements}).
The corresponding aerodynamic conditions ($s\simeq67$, $\mathrm{Ga}\simeq12755$) were thus similar to some of the ones by Greeley and Marshall~\cite{GreeleyMarshall85}. Upon increasing the wind free stream and thus shear velocity, which~we measured from extrapolating the log-layer law wind velocity profile (recorded using a hot-film anemometer), we determined $A$ and $A^\prime$, the values of which are also shown in Figure~\ref{ThresholdMeasurements}. It can be seen that, despite exhibiting the rather small dimensionless boundary layer thickness $\delta/(\sqrt{s}d)\simeq6.1$ (for which the comparable measurements by Greeley and Marshall~\cite{GreeleyMarshall85} exhibit large relative threshold gaps), there is nearly no relative gap between $A$ and $A^\prime$, which thus supports the derived dependency of $A/A^\prime$ on particle exposure to the flow [Equation~(\ref{TimeScaleRatio4})].

\section{Discussion and Conclusions} \label{Discussion}
\begin{figure}[H]
\centering
\includegraphics[width=0.98\columnwidth]{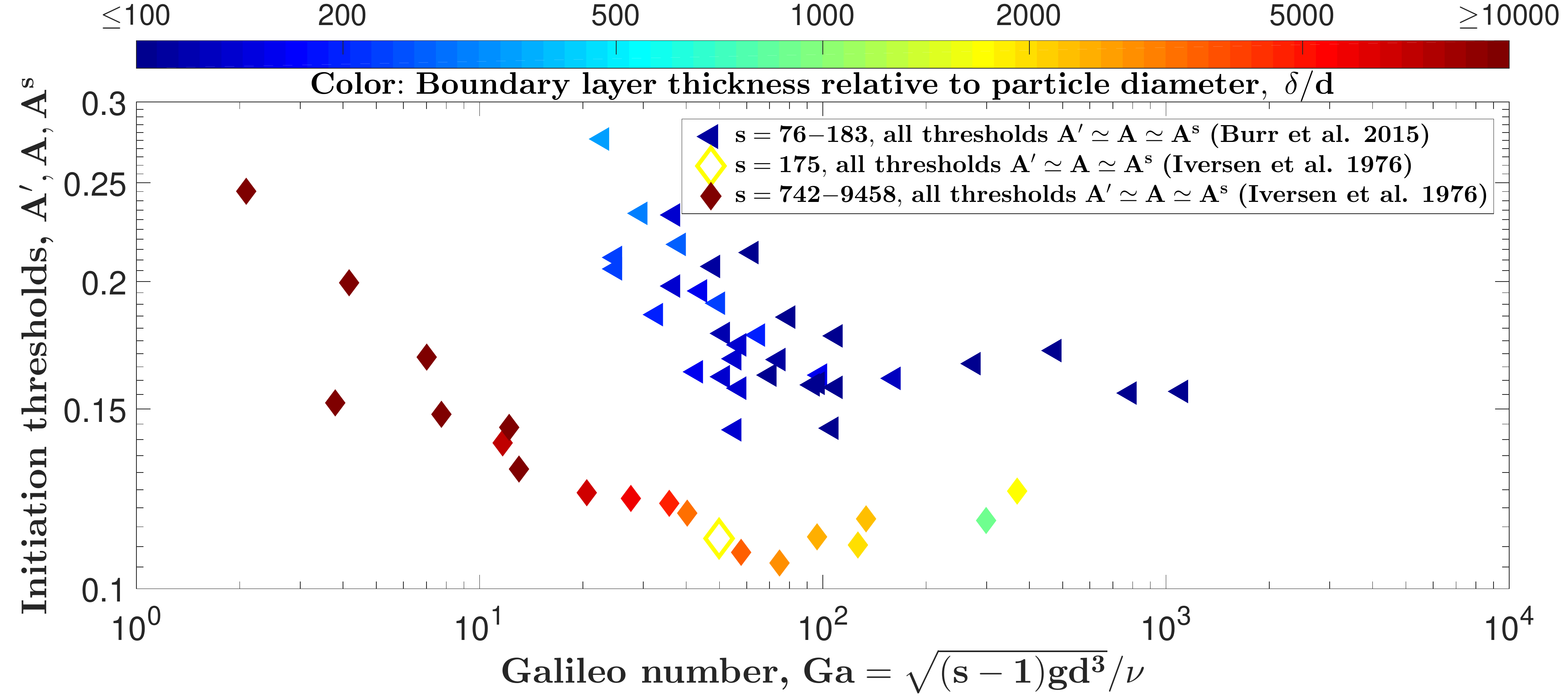}
\caption{Measurements of the rocking ($A^\prime$), rolling ($A$), and saltation ($A^s$) threshold parameters versus the Galileo number $\mathrm{Ga}$. Only those conditions from Figure~\ref{ThresholdMeasurements} that obey $A^\prime\simeq A\simeq A^s$~\cite{Iversenetal76,Burretal15} are shown. The color indicates the the thickness of the boundary layer $\delta$ relative to the particle diameter $d$ (different from Figure~\ref{ThresholdMeasurements}), which controls the relative amplitude of turbulent fluid velocity fluctuations for a constant $\mathrm{Ga}$.}
\label{ThresholdMeasurements2}
\end{figure}

Probably the most important insight provided by our study is that the current point of view~\cite{Iversenetal87,Burretal15} that the density ratio $s$ has a strong influence on the saltation initiation threshold $A^s$ for a constant Galileo number $\mathrm{Ga}$ (or a constant particle Reynolds number $\mathrm{Re}_d$) may not be true, at least in wind tunnels. This point of view has been established as a result of experiments carried out using the `Venus~Wind Tunnel'~\cite{Greeleyetal84}, also called `Titan Wind Tunnel'~\cite{Burretal15}, for conditions with intermediate density ratios ($s\sim100$). These experiments indicate significantly larger values of $A^s$ than one would expect from semi-empirical models calibrated for atmospheric conditions on Earth (e.g.,~\cite{IversenWhite82,Iversenetal87,ShaoLu00,Burretal15}). However, the Venus (Titan) Wind Tunnel produces a much thinner boundary layer ($\delta\approx1.9~$cm) than wind tunnels usually used to mimic atmospheric conditions on Earth and Mars ($\delta\sim1~$m), and~Figures~\ref{ThresholdMeasurements} and \ref{ThresholdMeasurements2} suggest that this may be the reason why the saltation thresholds ($A^s$) measured in this wind tunnel are larger, as we explain in the following.

First, one can see that the measurements of $A^s$ by Iversen et al.~\cite{Iversenetal76} (diamonds in Figures~\ref{ThresholdMeasurements} and~\ref{ThresholdMeasurements2}), obtained from experiments in a comparably large wind tunnel ($\delta\approx1.2~$m), follow a relatively smooth behavior as a function of $\mathrm{Ga}$ despite the fact that $s$ varies between $175$ and $9458$. In particular, the one measurement for $s\simeq175$ and $\mathrm{Ga}\simeq50$ (open diamond) does not deviate much from its two neighboring points ($(s,~\mathrm{Ga})\simeq(916,40)$ and $(s,~\mathrm{Ga})\simeq(2083,58)$), which one would expect if $s$ really had a significant influence on $A^s$ for a constant $\mathrm{Ga}$. Furthermore, this measurement by \mbox{Iversen et al.}~\cite{Iversenetal76} is significantly below the measurements of $A^s$ by Burr et al.~\cite{Burretal15} for similar values of $s$ and $\mathrm{Ga}$, which~also cannot be explained by the density--ratio hypothesis. However, this behavior is consistent with our theoretical framework (Figure~\ref{ThresholdMeasurements2}) because the relative magnitude of turbulent velocity fluctuations and thus the parameter $\alpha$ increase with $\delta/d$ for a constant $\mathrm{Ga}$ (Section~\ref{alpha}). Increasing $\alpha$ means that a weaker mean flow is required for aerodynamic entrainment (i.e., lower $A^\prime$ and thus $A^s)$. In other words, thresholds for a given sediment size can be lower when the boundary layer is relatively thicker. Hence, semi-empirical threshold models fitted to a certain wind tunnel data set may neither be applicable to other wind tunnels with significantly different boundary layer thickness $\delta$ nor to natural atmospheric conditions even when $s$ and $\mathrm{Ga}$ are the same.

Second, for the experiments by Burr et al.~\cite{Burretal15}, saltation and rocking thresholds become indistinguishable when the dimensionless boundary layer thickness $\delta/(\sqrt{s}d)\gtrsim6.6$ (Figure~\ref{ThresholdMeasurements}). However, when $\delta/(\sqrt{s}d)$ falls below this critical value, $A^s>A^\prime$, which is particularly apparent for the three bluish-colored measurements at $\mathrm{Ga}\approx600$ in Figure~\ref{ThresholdMeasurements}. These three measurements were largely responsible for the finding by Burr et al.~\cite{Burretal15} that $A^s$ is unusually large (see the three largest-diameter measurements in their Figure~2). However, for the natural atmosphere of Titan ($\delta\approx3~$km~\cite{Lorenzetal10}), $\delta/(\sqrt{s}d)$ is much larger, and our theoretical framework thus would predict that $A^s$ is very close to $A^\prime$ and thus considerably smaller than the values measured by Burr et al.~\cite{Burretal15} even if the additional effect of $\delta/d$ on $A^s$ from controlling the relative amplitude of turbulent velocity fluctuations (Figure~\ref{ThresholdMeasurements2}) was not considered. There is an easy test that should be done by future studies to figure out whether this prediction holds true. If the Venus (Titan) Wind Tunnel was used to measure $A^s$ for standard air (i.e., atmospheric conditions on Earth) rather than Titan-like air, the value $\delta/(\sqrt{s}d)$ would be even smaller because of a larger value of $s$, and one would thus expect an even larger deviation from semi-empirical threshold models when using the the Venus (Titan) Wind Tunnel with standard~air.

Finally, one must ask the question of whether wind tunnel experiments, in general, are~an appropriate means to estimate Aeolian sediment transport initiation thresholds under natural atmospheric conditions. Natural atmospheres exhibit much thicker boundary layers ($\delta\sim0.1{-}10~$km on Venus, Earth, Mars, and Titan~\cite{Lorenzetal10,Petrosyanetal11,Koketal12,Lebonnoisetal18}) and thus larger values of $\alpha$ (Section~\ref{alpha}), which are associated with smaller initiation thresholds (cf. Figure~\ref{ThresholdMeasurements2}). In fact, observations from the Mars rovers {\em{Opportunity}} and {\em{Curiosity}}~\cite{Silvestroetal13,Chojnackietal15,Bakeretal18} and the existence of gravel megaripples on Earth (e.g., in the Argentinean Puna~\cite{DeSilvaetal13}) suggest that the initiation of sediment transport may be much easier in the field than in wind tunnels. If this hypothesis were generally true, the problem of whether fully developed saltation can be initiated under natural atmospheric conditions would essentially reduce to the problem of whether saltation transport initiated by a large turbulent fluctuation event can be sustained by the mean turbulent flow. Answering this question requires knowledge of the dynamic saltation threshold $A^r$ (and the function $C(\mathrm{Ga},~s)$ discussed in Section~\ref{RollingSaltation}), for which turbulence only plays a minor role and for which an analytical model exists that reproduces available measurements in Aeolian and viscous and turbulent fluvial environments despite not being fitted to these measurements~\cite{PahtzDuran18a}. Hence, we would be able to predict saltation transport initiation on extraterrestrial planetary bodies much more reliably than we currently do. Another consequence would be that the saltation initiation threshold $A^s$ in the field, in contrast to the one in wind tunnels, would, like $A^r$, significantly decrease with the density ratio $s$ for a constant Galileo number $\mathrm{Ga}$~\cite{PahtzDuran18a}. Clearly, {\em{controlled}} field measurements are needed to test the hypothesis of whether sediment transport initiation in the field is truly much easier than in wind tunnels and thus to test whether all of these potential consequences hold true. Likewise, wind tunnel experiments in the spirit of Williams et al.~\cite{Williamsetal94} with the purpose to study the influences of the boundary layer thickness $\delta$ and turbulence frequency spectra on the different initiation thresholds and related quantities (e.g., bed particle removal rates, particle removal PDFs) can also shed more light on these questions.

\vspace{6pt} 

\authorcontributions{Conceptualization, T.P. and M.V.; Methodology, T.P. and M.V.; Validation, T.P., M.V. and X.-H.Z.; Formal Analysis, T.P.; Investigation, T.P., M.V. and X.-H.Z.; Resources, T.P., M.V. and Z.-S.L.; Data~Curation, T.P., M.V. and X.-H.Z.; Writing—Original Draft Preparation, T.P.; Writing—Review and Editing, T.P.~and M.V.; Visualization, T.P. and M.V.; Supervision, T.P., M.V. and Z.-S.L.; Project Administration, T.P., M.V.~and Z.-S.L.; Funding Acquisition, T.P. and M.V.}

\funding{This research was funded by the National Natural Science Foundation of China (Grant~Nos. 11750410687 and 41171005).}

\acknowledgments{We thank Devon Burr for providing us with the values of the rocking thresholds measured in the experiments by Burr et al.~\cite{Burretal15}.}

\conflictsofinterest{The authors declare no conflict of interest.} 

\reftitle{References}

\end{document}